# Defending the P-value

Yudi Pawitan
Department of Medical Epidemiology and Biostatistics
Karolinska Institutet, Stockholm, Sweden
yudi.pawitan@ki.se
original: 2020; revised: 8 January 2022

## Abstract

Attacks on the P-value are nothing new, but the recent attacks are increasingly more serious. They come from more mainstream sources, with widening targets such as a call to "retire" the significance testing altogether. While well meaning, I believe these attacks are nevertheless misdirected: Blaming the P-value for the naturally tentative trial-and-error process of scientific discoveries, and presuming that banning the P-value would make the process cleaner and less error-prone. However tentative, the skeptical scientists still have to form unambiguous opinions, proximately to move forward in their investigations and ultimately to present results to the wider community.  With obvious reasons, they constantly need to balance between the false-positive and false-negative errors. How would banning the P-value or significance tests help in this balancing act? It seems trite to say that this balance will always depend on the relative costs or the trade-off between the errors. These costs are highly context specific, varying by area of applications or by stage of investigation. A calibrated but tunable knob, such as that given by the P-value, is needed for controlling this balance. This paper presents detailed arguments in support of the P-value.

## Introduction

If you are an applied statistician or scientist, you are very likely using significance tests and producing P-values regularly. So it may not escape your attention that there are numerous attacks on the P-value. These attacks are of course not new, but the recent ones appear to come from mainstream sources and their target range has widened. The level varies from benign, to medium, to severe. On the benign end: Benjamin et al (2018) in *Nature Human Behaviour* simply suggested that we stop using 0.05 cutoff and start using 0.005 instead. Amrhein et al (2019) in *Nature* stepped up the attack and reported more than 800 supporting votes, including many in academia, in their call to "retire" statistical significance. They are not banning the P-value per se, but calling for the dismantling of significance testing. According to them, we should not say "statistically significant" and "not statistically significant"; in general, we should "quit categorizing".  They do not mind confidence intervals, but are of course against using them as the basis for claiming statistically significant results. "[A]uthors will emphasize their estimates and the uncertainty in them – for example, by explicitly discussing the lower and upper limits of their intervals."

Going further on the severity spectrum: the journal *Epidemiology* bans the P-value altogether, but not confidence intervals; actually this has been its practice since its founding in 1990. On the most extreme end of the spectrum, the journal *Basic and Applied Social Psychology* since 2015 had banned the P-value, or more generally the Null Hypothesis Significance Testing Procedure. They claim that this procedure is invalid, because it suffers from the "inverse inference problem." The confidence interval also suffers from the same problem, so it is also banned. For the Bayesian procedures, the Editors "reserve the right to make case-by-case judgements", so they are neither required nor banned.

The non-technical meta-literature on the P-value is enormous. Just the special issue of *The American Statistician* -- an official journal of the American Statistical Association -- in March 2019 alone contains 43 papers (401 total pages). It is practically impossible for a person to absorb the diversity of opinion among the experts, let alone the non-experts. What I find most interesting is perhaps the explicit opinion of the Editors, as to some extent they represent mainstream academic statisticians. Echoing Amrhein et al (2019), the Recommendation 2 in their editorial clearly states "Don't Say 'Statistically Significant', and they wrote: "… it is time to stop the term "statistically significant" entirely. Nor should variants such as 'significantly different', 'p < 0.05', and 'nonsignificant' survive."

This debate is not likely to die out soon. In August 2019 the *Journal of American Medical Association (JAMA) Psychiatry* published an article "Is It Time to Ban the P-value?" by HC Kraemer from the Department of Psychiatry and Behavioral Sciences, Stanford University, not exactly some out-of-the-way place. Book-length treatment is not as common. The title *The Cult of Statistical Significance* and the subtitle *How the Standard Error Costs Us Jobs, Justice, and Lives* of the book by Stephen Ziliak and Deirdre McCloskey (2008) clearly convey the views of the authors.

How are we to react to these attacks on the P-value and significance test? As you continue to read this article it should be become obvious which side I am on, so I might as well say it explicitly now: Banning P-value or significance test is a bad idea; "quit categorizing" defeats the purpose of scientific investigation; reducing the threshold from 0.05 or 0.005 might be sensible for new discoveries, but not a good idea as a blanket rule.

## When the P-value is unavoidable

There are a number of common statistical problems and applications where banning the P-value is awkward and not sensible. Here are some of them:

- Displaying and comparing Kaplan-Meier survival curves. These are among the most common plots in medical applications. What if we are not allowed to report the standard log-rank statistic and its P-value? We could report the hazard ratio and its confidence intervals, but this is model-dependent, for example it assumes proportional hazards. We could provide confidence bands, but then what do we state: are we allowed to focus on the areas where the confidence bands do not overlap? On the technical side, there is of course an issue of simultaneous inference across time axis, but this effort would also violate the idea of not using confidence intervals for significance testing.

- Wilcoxon and non-parametric rank tests in general. Technically there is an underlying parameter associated with a rank test, but who in practice knows what specific parameter is being tested by the Wilcoxon test? It is not at all clear whether an estimate and confidence interval for such a parameter, e.g. using the Hodges-Lehmann estimator, are informative for the readers.

- Trend tests across an ordinal axis; e.g. is there a trend in the risk of death across the low-medium-high exposure group? The test may depend on an arbitrary scaling of the x-axis, so the estimated slope and its confidence interval are not a meaningful numbers.

- Any test with more than one degree of freedom: ANOVA tests of multiple groups or interactions, or $\chi^2$ test of independence or goodness-of-fit or model-checking. When we test the association between, say, 4 eye colors and 3 hair colors, the $\chi^2$-statistic has (4-1) x (3-1) = 6 degrees of freedom. This means there are 6 independent underlying parameters that are being tested. What are they? It is possible to write them down, but they are simply not interesting. Providing the estimate and confidence interval for each would be absurd.

- Because of its importance in molecular medical research, in the genome-wide association studies, the plot of a very large collection (millions) of P-values – also called the Manhattan plot – is a primary output of the analysis. It is simply not sensible to show one million confidence intervals.

- In the ubiquitous multiplicity problems, the P-value is often the raw ingredient for further adjustment to account for multiplicity. It is very simple to adjust the P-value, for example, using the Bonferroni correction. If the estimates and confidence intervals are given, with some calculations, though not in your head, you can recompute the confidence intervals to adjust for multiplicity, but this is rarely done.

- From a theoretical point view, the construction of confidence interval requires a continuous parametric model, or at least semi-parametric, whereas the P-value only requires one model under the null hypothesis. For example, in spatial point-process applications we first want to test whether a homogeneous Poisson process fits the data. If yes, there is not much more we can investigate. It does not make sense to build a more complex nonhomogeneous Poisson model, for example, a competition process or a cluster point process, and to provide an estimate and confidence interval for the parameters of that process.

- Last but not least: a great contribution of statistics is in the theory of – or at least in the explicit considerations for -- design of experiments. Above all, thinking about design issues prior to starting the study forces a good discipline on the experimenters. Theoretically the concepts of testing, test statistic, significance level, power, and sample size are closely connected. Banning the P-value must also mean banning the test statistic, since there is one-to-one map between them. Without the test statistic it does not make sense to talk about power and sample size. Overall, much of the good statistical ideas from the theory of design of experiments would have to be abandoned.

Some journals that ban the P-value, such as *Epidemiology*, have in fact adopted some exceptions that cover some of those awkward cases, including the trend test, interaction test, or tests with multiple degrees of freedom.

## Reasons to ban the P-value

Let us assess some commonly stated reasons to ban the P-value. Number one, because people often misinterpret it as the probability of the null hypothesis. Is that really a mortal sin? P-value *is not* the probability of observed data given the null hypothesis (that would be the likelihood of the null). P-value is more complex: the probability of the observed data *or something more extreme than the observed data* assuming the null hypothesis is true. So P-value is not a transversal of the Bayesian posterior:

> P-value = P(observed data or more extreme|$H_0$)
> Bayesian posterior = P($H_0^*$|observed data).

(The notation $H_0^*$ is used for the Bayesian posterior to indicate it may not be exactly the same set as $H_0$ for the P-value. E.g. $H_0$: $\mu = 0$, but $H_0^*$: $\mu \leq 0$ for a typical testing of the normal mean parameter.)

In regular problems, the kind that lead to complaints about the inferential meaning of P-value, the P-value is indeed not a probability of the null hypothesis, but *it is* a valid measure of "confidence" (Pawitan 2001, Chapter 5). This non-Bayesian concept of confidence is the same "confidence" that we use in "confidence interval," so it is a fully mainstream concept. Unfortunately, it is rarely used in its full capacity as a measure of uncertainty. The recent book by Schweder and Hjort (2016) gives a comprehensive exposition. Thus, we may report, say, the 95% confidence interval $2.1 < \mu < 3.9$, meaning that specific interval has 95% confidence. The (one-sided right-side) P-value for the null hypothesis $H_0$: $\mu = 0$ is the *confidence associated with $\mu \leq 0$*.

Furthermore, theoretically, the confidence distribution is a Bayesian posterior for a certain implied prior distribution (implied by the choice of P-value) (Fraser, 2011; Efron and Tibshirani, 1993, Chapter 24). In summary, mathematically we may interpret the P-value as a confidence measure and as a Bayesian posterior probability. So the reason to ban the P-value because it is not a probability of null hypothesis is not theoretically justified. It is a confidence measure in a relevant null hypothesis (such as $\mu \leq 0$ for one sided P-value) and it has the same logical status and epistemic value as the Bayesian posterior probability for a certain choice prior distribution. Now we can see also that to ban the P-value, but "might consider" the Bayesian alternative, shows a lack of theoretical understanding.

The second common reason: people use the P-value carelessly, thus produce many misleading results, primarily false positives or hyped claims. We know this because many results indeed fail to replicate. Statisticians should of course care about not producing misleading results. Most of our theories in statistics is based on optimizing error rates. But, as we all know, there is a type-I error and a type-II error. Whatever we say about a hypothesis, there is a chance we make an error. If we declare significance we may make type-I error, which is a sin of commission; if non-significance, we might commit type-II error, a sin of omission. A good procedure should balance between the two errors, and a good statistician or scientist should be aware of both at all times. But how is that achieved by banning the P-value?

Benjamin et al (2018)'s modest suggestion of reducing the P-value threshold of significance to 0.005 goes a long way towards reducing false positive rates. But even this sensible suggestion cannot be applied as a blanket rule to all studies, because it will increase the type-II error rate. For example, for a validation study, a 0.05 threshold could be sufficient. Together with the original study, the cumulative evidence for real effect is much stronger. But, in practical scientific activities, this is where a grey area appears: what constitutes a "validation" study? Does it have to be by the same research group as the original discovery? If not, does it count as validation if the idea of the study comes after reading a journal article?

But scientists might argue that most of their studies are based on a chain of previous studies and results, either by themselves or by their collaborators or by complete strangers; the studies are never completely new, based on an idea that appears out of the blue. Also, to be a "validation" study, does the study have to be as similar as the original? The original study could be done in mice, does a human version constitute a validation? A phase-III clinical trial to get the US Federal Drug Agency's approval for a drug typically use 0.05 threshold for significance. At that point the drug has already undergone thorough and numerous investigations to establish its biological basis, so the human trial is a form of validation.

Eventually, if a question is important enough, there should be many similar studies performed as (near)-replicates, so a systematic meta-analyses can be done to get a more definitive answer with a very strong P-value.

Perhaps the most legitimate concern about the P-value is that it does not capture the increased uncertainty due to various steps and decisions – some documented and some undocumented -- we make during data analysis: choice of wording in a questionnaire, definition of new variables for analysis, choice of transformation, choice of split when analyzing categories, choice of model class, decisions regarding inclusion and exclusion criteria, choice of variable(s) for sub-group analysis, handling of missing data, etc. Explicit model selection, for example, the best-subset regression, is known to produce optimistic P-values from the final selected model. But how do we account for all various steps, some of which are often taken with the thought of increasing the sensitivity of the experiment, hence potentially giving an optimistic P-value?

Overcoming the weakness of the P-value as a discovery tool, science has a safety measure by evolving a skeptical culture. It is rare for any scientific result to be readily accepted at first sight by the relevant research community – the group most knowledgeable with the details of the area, hence not as easily fooled as the general public. The simple requirement is that a result gets established by external validation and reproducibility. The harder the validation the stronger the result would be. What constitute a "hard validation"? As I have indicated above, a scientific validation does not mean a result based on a study as similar as possible to the original study. In fact, it is better to validate in distinct studies. For example, a biological finding from human study can be first validated in zebra-fish, then in mice, then back in human. All relates to the same biology, but perhaps with different measurements. Good scientific results should lead to a deeper theory, which in turn should produce novel predictions for future experiments. These future experiments, even with completely distinct measurements, will constitute a validation of the original study or theory. In summary, reproducibility is a must in science, though it must be seen as a general concept.

Most opponents of the P-value and its 0.05 threshold point their finger to R.A. Fisher, one of the greatest statisticians of the 20th century. With his highly influential *Statistical Methods for Research Workers*, including the practical Tables for various common statistics, Fisher certainly popularized the 0.05 limit. In a paper in 1926 he wrote (with my emphasis):

> … Personally, the writer prefers to set a <u>low standard of significance at the **5 per cent**</u> point, and ignore entirely all results which fail to reach this level. <u>A scientific fact should be regarded as experimentally established only if a properly designed experiment **rarely fails** to give this level of significance.</u>

As a practicing scientist, Fisher started with a "low standard" 0.05 limit, but note the key words "*rarely fail*" as it clearly implies that he required successful validation studies before accepting a scientific fact as established. He was also careful to state that this was *his personal preference*. The proper balance between type-I and type-II error is highly context-dependent. In research areas where discoveries are rare, or, if a potential discovery – if real – is fundamental, or, if false positives are easy to dismiss, it is not sensible to have a very stringent P-value cut-off, as it lowers sensitivity. In such a situation the researchers would be happy to pursue any reasonable lead. Though Fisher did not elaborate, he obviously did not mind the cost of false positives from his "low standard" threshold, since it gave him higher chances of seeing the true positives.

On the other hand, when potential discoveries are abundant *and* establishing a scientific fact is expensive – for example, in genome-wide association studies (GWAS) or in high-energy physics – one should be very stringent with P-value cutoff. The standard cutoff is $5 \times 10^{-8}$ in GWAS, and $3 \times 10^{-7}$ in high-energy physics.

## How to quit categorizing?

I believe the proposal to ban the term "statistically significant" or "significantly different", and to "quit categorizing" is downright unhelpful. Unfortunately, the proposers do not give good examples how to report statistical results. So let us try this ourselves. Mack et al (2019) reported the result of the study on transcatheter aortic-valve replacement with a balloon-expandable valve in the *New England Journal of Medicine*. Here is how the result and conclusion paragraphs appear in the Abstract. Some offending parts are highlighted with bold or underline, and both if we quit categorizing and ban the P-values; they do cover much substance of the results:

> RESULTS At 71 centers, 1000 patients underwent randomization. The mean age of the patients was 73 years, and the mean Society of Thoracic Surgeons risk score was 1.9%

> (with scores ranging from 0 to 100% and higher scores indicating a greater risk of death within 30 days after the procedure). The Kaplan–Meier estimate of the rate of the primary composite end point at 1 year was **significantly lower** in the TAVR group than in the surgery group (8.5% vs. 15.1%; absolute difference, −6.6 percentage points; 95% confidence interval [CI], −10.8 to −2.5**; P<0.001 for noninferiority**; hazard ratio, 0.54; 95% CI, 0.37 to 0.79; **P=0.001 for superiority**). At 30 days, TAVR resulted in a **lower rate of stroke than surgery (P=0.02)** and in **lower rates of death or stroke (P=0.01)** and **new-onset atrial fibrillation (P<0.001).** TAVR also resulted **in a shorter index hospitalization than surgery (P<0.001)** and in a **lower risk of a poor treatment outcome (death or a low Kansas City Cardiomyopathy Questionnaire score) at 30 days (P<0.001)**. There **were no significant between-group differences** in major vascular complications, new permanent pacemaker insertions, or moderate or severe paravalvular regurgitation."
>
> CONCLUSIONS Among patients with severe aortic stenosis who were at low surgical risk, the rate of the composite of death, stroke, or rehospitalization at 1 year was **significantly lower** with TAVR than with surgery.

Here is what happens if we clean the Abstract from the offending parts, still allowing confidence intervals, but not P-values:

> At 71 centers, 1000 patients underwent randomization. The mean age of the patients was 73 years, and the mean Society of Thoracic Surgeons risk score was 1.9% (with scores ranging from 0 to 100% and higher scores indicating a greater risk of death within 30 days after the procedure). The Kaplan–Meier estimate of the rate of the primary composite end point at 1 year was 8.5% in the TAVR group and 15.1% in the surgery group (absolute difference, −6.6 percentage points; 95% confidence interval [CI], −10.8 to −2.5; hazard ratio, 0.54; 95% CI, 0.37 to 0.79).

As a scientific statement, the last sentence looks acceptable, but, without categorical direction, the statement seems dry and harder to absorb than the original. The second half of the original abstract must go because all statements are categorical. Finally, as a plea to all proposers of banning the significance test and the term "significantly different," please tell us how to formulate the conclusion statement. If "significantly lower" is not allowed, how about just "lower"? But it is still categorical, and it will be less informative to the readers. The second half of the Results paragraph looks like "sizeless science" as complained by Ziliak and McCloskey (2008), but the results could of course be reported in terms of statistics and confidence intervals. They would just entail an undigested list of numbers as given in Table 2 of the paper (below). Thus effect sizes of the secondary end-points are indeed available within the paper, but the authors have to make choices due to limited space in the Abstract. Even stylistic considerations alone would stop most authors from simply giving a tedious list of numbers in their Abstract.

| End Point | TAVR (N=496) | Surgery (N=454) | TAVR vs. Surgery (95% CI)† | P Value‡ |
|---|---|---|---|---|
| New-onset atrial fibrillation at 30 days — no./total no. (%)§¶ | 21/417 (5.0) | 145/369 (39.5) | 0.10 (0.06 to 0.16) | <0.001 |
| Length of index hospitalization — median no. of days (interquartile range) | 3.0 (2.0 to 3.0) | 7.0 (6.0 to 8.0) | −4.0 (−4.0 to −3.0) | <0.001 |
| Death from any cause, stroke, or rehospitalization at 1 year — no. (%)§ | 42 (8.5) | 68 (15.1) | 0.54 (0.37 to 0.79) | 0.001 |
| Death, KCCQ score of <45, or decrease from baseline in KCCQ score of ≥10 points at 30 days — no./total no. (%)‖ | 19/492 (3.9) | 133/435 (30.6) | −26.7 (−31.4 to −22.1) | <0.001 |
| Death or stroke at 30 days — no. (%)§ | 5 (1.0) | 15 (3.3) | 0.30 (0.11 to 0.83) | 0.01 |
| Stroke at 30 days — no. (%)§ | 3 (0.6) | 11 (2.4) | 0.25 (0.07 to 0.88) | 0.02 |

Table 2 from Mack et al (2019), *New England Journal of Medicine* 380: 1695-1705.

## Life without any statistical inference

How is life without the P-value or any statistical inference in the journal Basic and Applied Social Psychology (BASP)? Here is one example from a recent article: "Sorry is the Hardest Word to Say: The Role of Self-Control in Apologizing" by Guilfoyle et al (2019). The Abstract is as follows; the key statistical results are highlighted.

> "Apologizing is an effective strategy for reconciling relationships after transgressions. However, transgressors often resist or refuse to apologize. The current research investigated the role of self-control in apologizing. <u>In Study 1, self-control was associated with participants' proclivity to apologize and apologetic and nonapologetic behavior</u>. In Studies 2 and 3, self-control was manipulated to test the causal relationship. **Both studies found participants with high self-control were more apologetic and less nonapologetic and were more likely to use apologetic statements in e-mails to their victims**. <u>Overall, these studies suggest that transgressors with high self-control are more apologetic than those with low self-control</u>."

Now, in the body of the paper, we can find the supporting evidence for the statements in the Abstract:

> …, those in the high self-control group reported greater apology ($M = 5.08$, $SD = 1.45$) than those in the low self-control group ($M = 4.76$, $SD = 1.50$), $d = 0.22$ (Figure 1). We also found that those in the low self-control condition reported greater nonapology ($M = 3.12$, $SD = 0.79$) than those in the high self-control group ($M = 2.95$, $SD = 1.00$), $d = 0.19$ (Figure 2).

Hence, as promised, no statistical tests need to be reported, though it is not clear if the authors had actually performed them prior to submission to BASP. Interestingly, the text does not state the standard errors, but the figures do. Using the standard deviations (SDs) in the figures would give alarmingly large error bars; but it is arguable whether you should be allowed to show standard errors, since they are inferential quantities, not summary statistics of the data. Is this kind of scientific reporting meant to be better than reporting with P-values? Remarkably, the impact factor of BASP increased from 1.3 in 2015 to 3.4 in 2017, no doubt giving the editors a great feeling of vindication, but it then went down to 1.0 in 2018 and 1.58 in 2019.

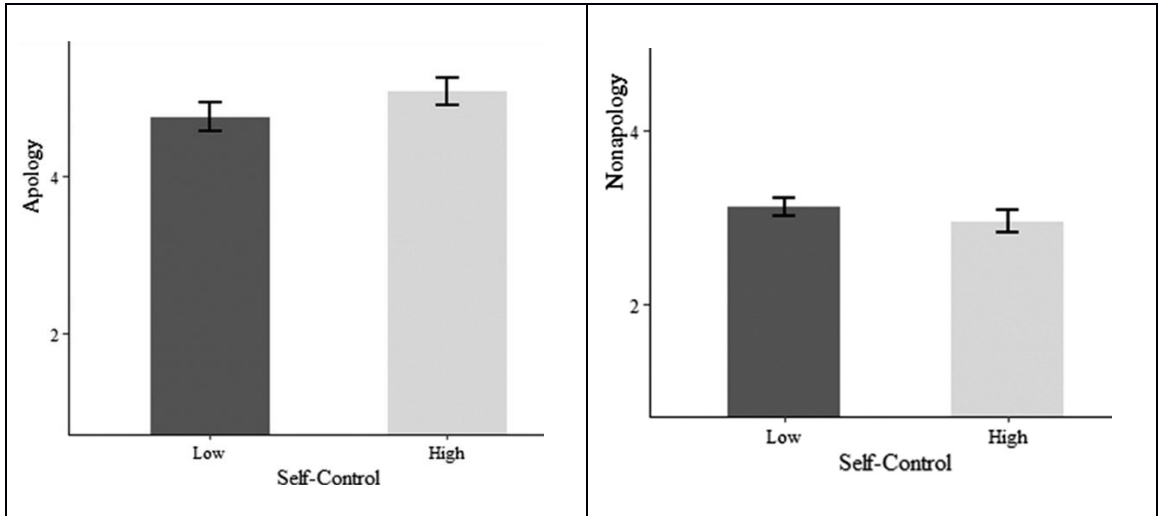

Figures 1 and 2 from Guilfoyle et al (2019), *Basic and Applied Social Psychology* 41: 72-90. The error bars represent standard errors.

If you are like me, you are probably curious as to what kind of people want to ban the P-value or its threshold or statistical inference. Among the 43 papers in the 2019 Special Issue of TAS, there is one by Valentin Amrhein, David Trafimow and Sander Greenland. Now S Greenland is well known to many statisticians; V Amrhein is the one who wanted to ban statistical significance; and D Trafimow is the Editor of BASP. The article is titled "Inferential statistics as descriptive statistics: there is no replication crisis if we don't expect replication." Here is the Abstract, which makes for strange reading for a scientist, and I highlight one sentence that perhaps captures the attitude of the authors:

> "Nonetheless, considerable non-replication is to be expected even without selective reporting, and generalizations from single studies are rarely if ever warranted. Honestly reported results *must* vary from replication to replication because of varying assumption violations and random variation; excessive agreement itself would suggest deeper problems. … Since a small P-value could be large in a replication study, and a large P-value could be small, **there is simply no need to selectively report studies based on statistical results.** Rather than focusing our study reports on uncertain conclusions, we should thus focus on describing accurately how the study was conducted, what problems occurred, what data were obtained, what analysis methods were used and why, and what output those methods produced."

In the text, we can read

> "Reaching for statistical tests to force out "inferences" (whether traditional "p≤α" testing or substitutes like tests using Bayes-factor criteria) is, like drinking alcohol, a culturally ingrained habit. Statistical testing (like alcohol) often gives the wrong impression that complex decisions can be oversimplified without negative consequences.... And many researchers are addicted to such oversimplification. These addictions are worth breaking."

## Conclusions

I guess that the people who write proposals to ban the P-value or significance tests, or the editors of journals that ban them, are well-meaning with no hidden agenda. I cannot see any profit to be made by holding such a point of view. However, we know that it is possible to be well-meaning and unhelpful. There is a tendency in their writing to consider only the common problem of comparing two groups with simple parametric one-degree of freedom tests; I mentioned several common problems where it is awkward not to use the P-value. The unfortunate result is that good suggestions – e.g. to pay attention to the estimates of effect sizes – are mixed with bad proposals and arguments. For example, Ziliak and McCloskey (2008) invented the term "sizeless science" as the straw man in their polemical book. In practice, when results are written down for publications, other factors come into play. As we show above, in the Abstract section of a paper, limited space considerations might lead to few details about effect sizes. But it would be highly unusual for the Results section not to report effect sizes.

In the process of developing the light-bulb, between 1878 and 1880 the legendary inventor Thomas Alva Edison (and his associates) famously worked on at least 3,000 different theories, trying out no fewer than 6,000 materials. Let us imagine his state of mind during the period: would he be arbitrary, not paying attention to the signal-to-noise ratio? How stringent would he be in the beginning investigation of a new material or design? How stringent would he be when a filament is showing promise? Finally, how stringent would he be on the design that he would be happy to patent? I imagine, like Fisher, most scientists today would work in a similar way: in the beginning quite happy looking for many possible leads, not afraid of false positives, but not going in completely arbitrary direction either; and, once there is a promising direction, putting a lot of efforts to validation studies.

The biggest difference is of course that Edison did not have to publish anything. Modern scientists have to publish even not-so-final findings; once published, there is no control who would consume the information and how the information is further disseminated. The skeptical scientists have no problem with the tentative nature of scientific results, even when presented categorically, as they would expect that validation studies will eventually screen out false positives. But the general public might get confused with this potentially chaotic process; this issue of course deserves its own discussion session. The point I would like to make here is that banning the P-value and significance tests seems to be motivated by the wish to reduce/avoid public confusion of scientific results, which are tentative by their nature, but this wish is contrary to the daily activities and needs of practicing scientists.

Finally, as is well-known in statistics we have a long-standing unresolved philosophical debate between the frequentist and the Bayesian schools. Most Bayesians do not accept the P-value; theoretically the P-value violates the likelihood principle (Pawitan, 2001, Chapter 7). And, they also claim that the P-value misrepresents the level of evidence against the null hypothesis. They would instead suggest the Bayes Factor as an alternative to the P-value. So, an attack on the P-value from the Bayesians is nothing new. It is possible that some of the current attackers are partly motivated by the old Bayesian dislike of the P-value, but the proposal to ban significance tests altogether should also apply to the use Bayes Factor, so it appears distinct from the Bayesian-frequentist debate.

## Acknowledgment

This manuscript was originally published in Qvintensen (Nr 2 2019), a bulletin of the Swedish Statistical Society. I am grateful to the Editor, Jan Wretman, who arranged a series of papers on the P-value in the bulletin and kindly helped in editing the manuscript. I am also grateful to my dear colleagues at MEB, particularly to Cecilia Lundholm, who first suggested I write this manuscript, and to Arvid Sjölander and Erin Gabriel for useful comments.